\documentclass[twoside]{dis08}
\usepackage[latin1]{inputenc}
\usepackage[dvips]{graphicx,epsfig,color}
\usepackage{wrapfig,rotating}
\usepackage{amssymb,amsmath,array}

\begin{document}
\title{Searches for Contact Interactions at HERA}

\author{Aleksander Filip \.Zarnecki \\
  (on behalf of the H1 and ZEUS collaborations)
%
%
\vspace{.3cm}\\
%
 Institute of Experimental Physics, University of Warsaw \\
   Ho\.za 69, 00-681 Warszawa, Poland
%
}

\maketitle

\begin{abstract}
The H1 and ZEUS collaborations at HERA have searched 
for signatures of physics beyond the Standard Model 
with  high $Q^2$ neutral current deep inelastic 
electron-proton and positron-proton scattering events.
No significant deviations 
from Standard Model predictions were observed. 
Various $eeqq$ contact interaction models 
have been considered.
Limits on  the compositeness scale in general $eeqq$ contact 
interaction models,  mass to the Yukawa coupling ratio for heavy leptoquarks,
the effective Planck mass scale in models with large extra dimensions
and the effective quark charge radius are presented. 
\end{abstract}


\section{Introduction}

The H1 and ZEUS experiments at HERA (DESY, Hamburg) allowed the study 
of electron-proton and positron-proton collisions at center
of mass energies of up to 920 GeV.
During the so called HERA~I running phase (1994-2000) about 100 $pb^{-1}$
of data were collected per experiment, mainly coming from $e^+ p$ collisions.
After the collider upgrade in 2000-2001, resulting in a significant
increase of luminosity, about 400 $pb^{-1}$ of data  per experiment
were collected in the so called HERA~II phase (2002-2007).
Moreover, spin rotators installed at the H1 and ZEUS interaction
regions provided longitudinal electron and positron polarization.
With average lepton beam polarization of about 30-40\% and significant
increase of integrated data luminosity (especially of collected 
$e^- p$ sample) HERA~II has significantly increased the sensitivity 
of the experiments to physics beyond the SM.


\section{Contact Interactions}

New interactions between electrons and quarks involving mass scales 
above the center-of-mass energy can modify the  deep inelastic $e^\pm
p$ scattering cross sections at high $Q^2$ via virtual effects,
resulting in observable deviations from the Standard Model predictions.
Four-fermion contact interactions are an effective theory, which 
allows us to describe such effects in the most general way.
Vector $eeqq$ contact interactions considered at HERA
can be represented 
as an additional terms in the Standard Model Lagrangian:
\begin{eqnarray*}
L_{CI} & = & \sum_{i,j=L,R} \eta^{eq}_{ij} (\bar{e}_{i} \gamma^{\mu} e_{i} )
              (\bar{q}_{j} \gamma_{\mu} q_{j}) 
\end{eqnarray*}
where the sum runs over electron and quark helicities,
and a set of couplings  $\eta^{eq}_{ij}$ describe 
the helicity and flavor structure of contact interactions.
%
%
Various scenarios, with different chiral structures, were considered by
H1 and ZEUS.  
Limits on the model parameters were derived from the analysis of
neutral current deep inelastic scattering (NC DIS) events.
Details of the analyses are described in~\cite{h1ci,zeusci}.


\section{Results}

\begin{figure}[htb]
\centerline{
\includegraphics[width=0.55\columnwidth]{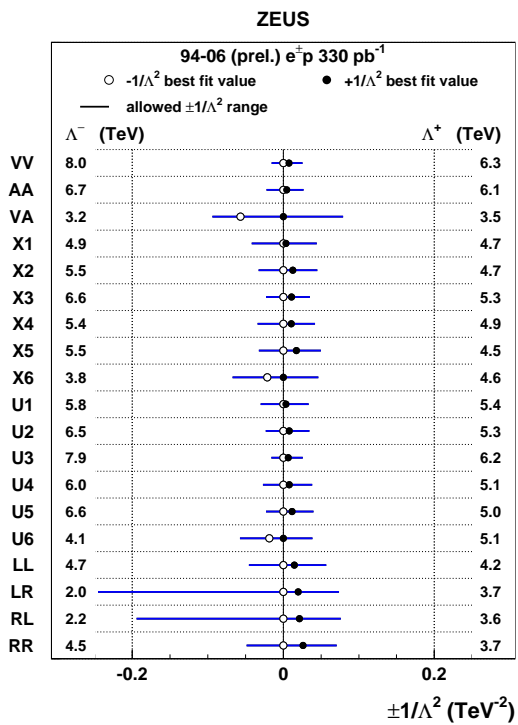}}
\caption{Results for general contact interaction models
  (compositeness models) obtained using the combined  $e^{+}p$ and $e^{-}p$ 
  data from ZEUS (1994-2006). Horizontal bars indicate the 95\% CL limits on
  $\eta / 4\pi = \varepsilon / \Lambda^{2}$; 
  values outside these regions are excluded.
  $\Lambda^{\pm}$ are the  95\% CL limits on the compositeness scale
  for $\varepsilon = \pm 1$.}\label{Fig:CI}
\end{figure}

In the general case (also referred to as compositeness models), 
limits on the effective ``new physics'' mass scale $\Lambda$ 
(compositeness scale) are extracted assuming the relation 
$\eta = \pm 4\pi / \Lambda^{2}$.
Figure~\ref{Fig:CI} shows the results obtained by the ZEUS 
for different compositeness models, based on the analysis of
1994-2006 data.
Limits on the effective mass scale $\Lambda$
range form 2 up to 8 TeV.
Corresponding limits obtained by the H1 collaboration, based
on the HERA I data only, range from 1.6 to 5.5 TeV~\cite{h1ci}.

\begin{figure}[p]
\centerline{
\includegraphics[width=0.57\columnwidth]{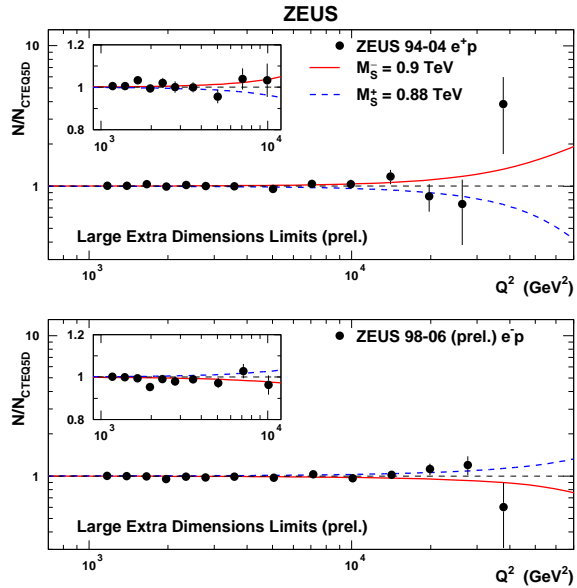}}
\caption{ZEUS data compared with 95\% CL exclusion limits for 
  the effective Planck mass scale in models with large extra 
  dimensions, for positive ($M_{S}^{+}$) and negative  ($M_{S}^{-}$) 
  couplings. Results of the experiments are normalized to the
  Standard Model expectations using CTEQ5D parton distributions.}\label{Fig:LED}
\end{figure}

For the model with large extra dimensions~\cite{add} both
collaboration set limits on the effective Planck mass scale $M_{S}$.
For negative coupling sign scales below
0.90~TeV (ZEUS 1994-2006) and 0.78~TeV (H1 1994-2000~\cite{h1ci}) 
are excluded on 95\% CL. For positive couplings the limits
are 0.88~TeV and 0.82~TeV respectively.
Possible effects of graviton exchange on the $Q^{2}$ distribution
of NC DIS events, as measured by ZEUS, are shown in Figure~\ref{Fig:LED}.

 \begin{figure}[p]
\centerline{
\includegraphics[width=0.6\columnwidth]{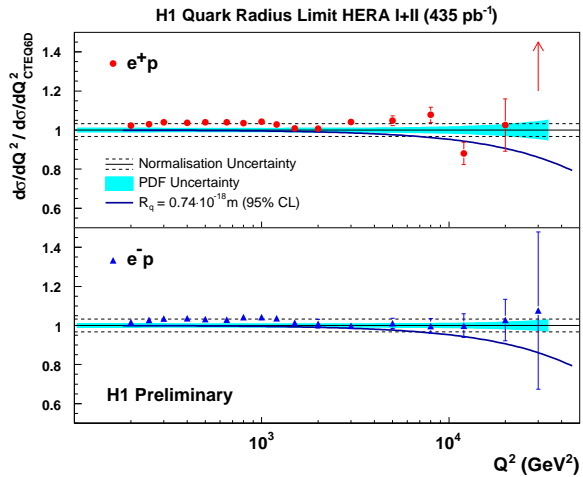}}
\caption{H1 data compared with 95\% CL exclusion limits for 
  the effective radius of the quark~\cite{h1rq}. 
Results of the experiments are normalized to the
  Standard Model expectations using CTEQ6D parton distributions.}\label{Fig:Rq}
\end{figure}

 \begin{figure}[htb]
\centerline{
\includegraphics[width=0.44\columnwidth]{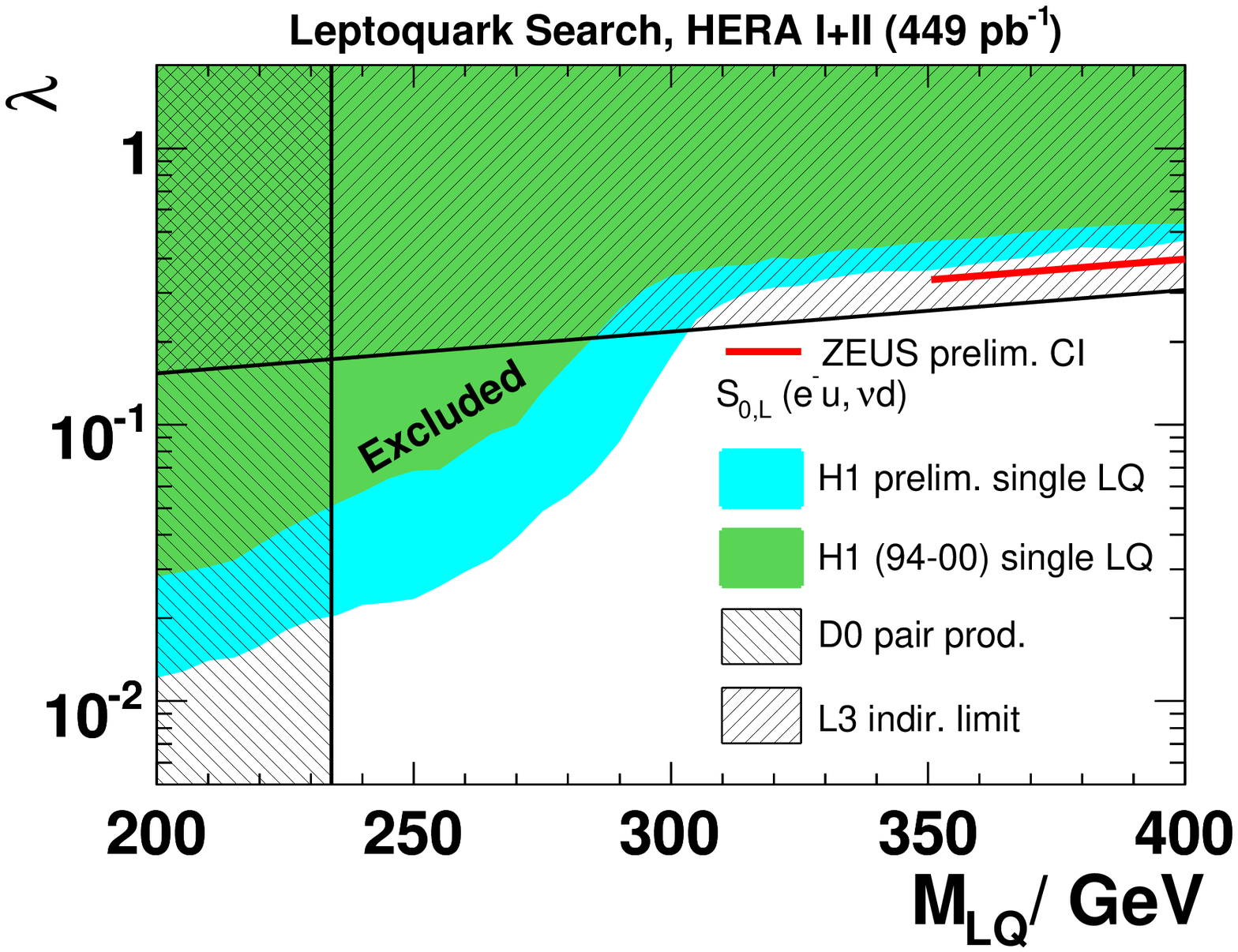}
\includegraphics[width=0.44\columnwidth]{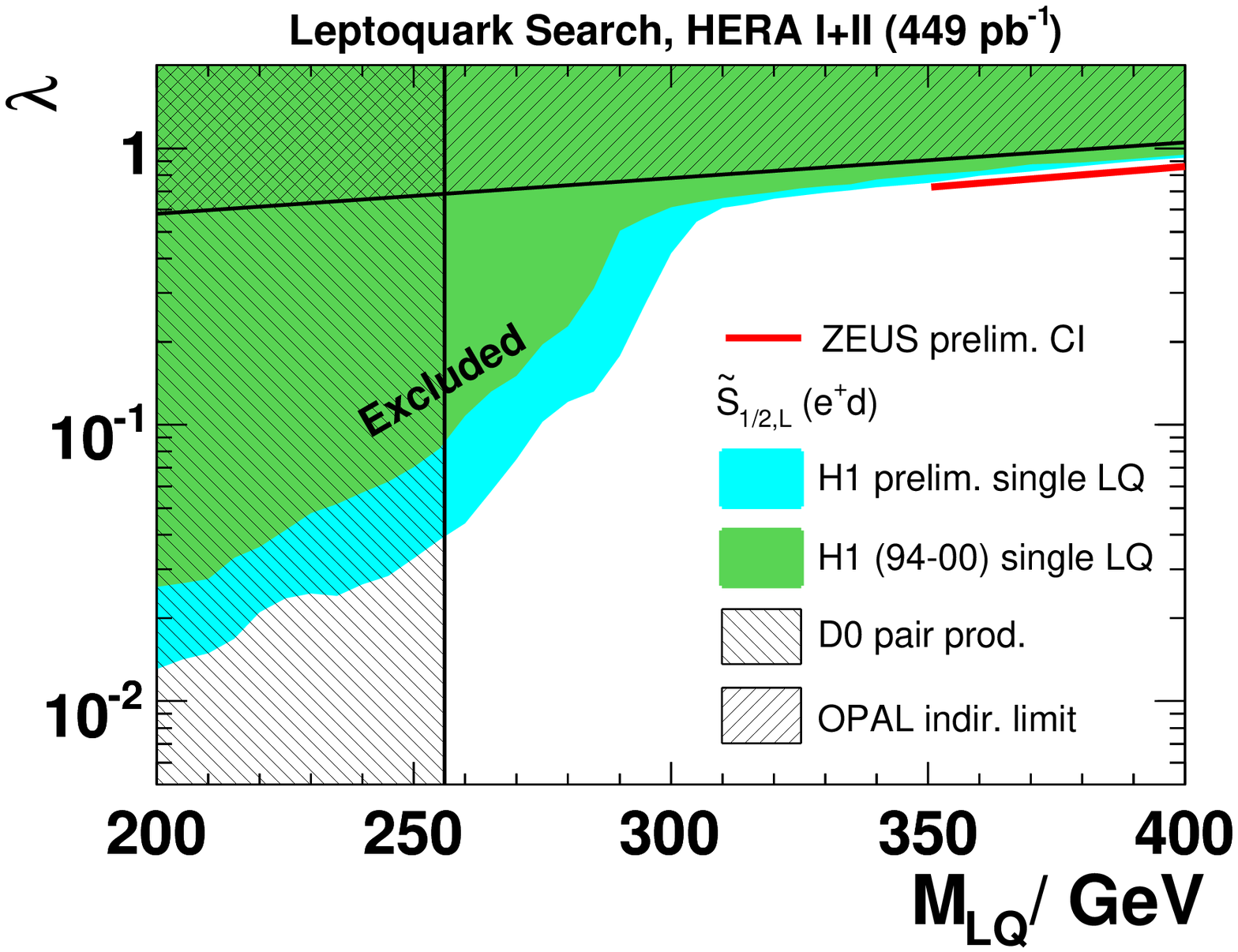}}
\caption{H1 exclusion limits at 95\% C.L. on the coupling 
 as a function of the leptoquark mass for $S_{0,L}$ 
and  $\tilde{S}_{1/2,L}$ leptoquarks~\cite{h1lq}. 
The indirect limits from ZEUS and L3 and the direct D0 limits are shown 
for comparison. }\label{Fig:LQ}
\end{figure}

Also, searches for possible quark substructure can be made by  measuring
the spatial distribution of the quark charge.
By using the ``classical'' form factor approximation, and assuming that
both electron and exchanged bosons are point-like, limits on the
mean-square radius of the electroweak charge of the quark can be set.
From the analysis of combined HERA I and HERA II  data quark radii
bigger than $0.74\cdot 10^{-16}$~cm (H1) 
and $0.62\cdot 10^{-16}$~cm (ZEUS) have been excluded at 95\% CL.
Figure~\ref{Fig:Rq} shows the H1 data together with 
95\% C.L. exclusion limits for the effective 
radius of the quark~\cite{h1rq}.

Contact interactions can also be used to describe the effects of
virtual leptoquark production or exchange at HERA, in the limit of
large leptoquark mass $M_{LQ} \gg \sqrt{s}$.
The ZEUS collaboration have used data taken between 1994 and 2006 to constrain
the leptoquark Yukawa coupling for different leptoquark types and masses.
Limits on the ratio between mass and Yukawa coupling range from 
0.29 to 2.08~TeV.
In Figure~\ref{Fig:LQ} indirect limits from ZEUS are compared to 
the H1 exclusion limits at 95\% C.L. on the Yukawa coupling of $S_{0,L}$
and $\tilde{S}_{1/2,L}$ leptoquarks, as a function 
of their mass~\cite{h1lq}.
Limits from LEP and the Tevatron are also indicated.

\section{Conclusions}

Lepton beam polarization and high luminosity delivered at HERA II
opened a new window for precise EW studies and searches for physics
beyond SM.
Measured NC DIS cross sections at high $Q^2$ are in very good 
agreement with SM, so only limits on deviations from SM could be set 
within different models.
HERA running has finished, but analyses of large samples of data
are still ongoing and more interesting results can be expected.


\begin{footnotesize}

\end{footnotesize}


\end{document}